\documentclass[a4paper,10pt,twoside]{cpc-hepnp}

\usepackage{multicol}
\usepackage{graphicx}
\usepackage{booktabs}
\usepackage{amssymb,bm,mathrsfs,bbm,amscd}
\usepackage[tbtags]{amsmath}
\usepackage{lastpage}
\usepackage{CJK}
\usepackage{natbib}

\begin{document}

\begin{CJK*}{GBK}{song}



\title{Test of the prototype of electron detector for LHAASO project using cosmic rays\thanks{Supported by NSFC (11075095) and Shandong Province Science Foundation (ZR2010AM015) }}

\author{%
      XU Tongye$^{1}$
\quad DU Yanyan$^{1}$
\quad SHAO Ruobin$^{1}$\\
\quad WANG Xu$^{1}$
\quad ZHU Chengguang$^{1;1)}$\email{zhucg@sdu.edu.cn} for LHAASO collaboration%
}
\maketitle

\address{%
$^1$ MOE key lab on particle physics and particle irradiation,\\Shandong University, Ji'nan 250100, China
}

\begin{abstract}
LHAASO project is to be built in south-west China, which use an array of 5137 election detectors for the measurement of the incident electrons arriving at the detector plane. For the quality control of the big quantity of electron detectors, a cosmic ray hodoscope with two-dimensional spacial sensitivity and good time resolution has been developed. The first prototype of electron detector is tested with the hodoscope and the performance of the detector is validated to be consistent with the design.
\end{abstract}




\begin{multicols}{2}
\section{Introduction}
The Large High Altitude Atmosphere Shower Observation (LHAASO)~\citep{Cao2010} is about to be constructed in south west China. An array consists of 5137 electron detectors (ED) is an essential part of this experiment, which are used to count the number of incident electrons arriving at the detector plane and to measure the incident angle of the original cosmic particles arriving at the earth. To achieve the required measurement precision of the incident angle of original cosmic particle, the criteria of the time resolution of the ED is set to be less than 2 nano-seconds. As the ED is a large area (1 square meter) scintillation detector, the uniformity of the responses, such as detection efficiency, photon yield, time measurement over all the sensitive area to single incident particle is very important for the precision of measurements and for the problem finding, which should be tested.

To achieve an efficient and robust test of the big quantity of EDs for the purpose of quality control, in a limit duration of time, a large COmic RAy Reference System (CORARS) making use of the cosmic muons arriving at the laboratory was developed. By design, the system provides more precise information of time and position of the arriving cosmic muons than ED as reference for the measurement listed above. The time resolution, energy spectrum and detection efficiency of of single particle of eight EDs can be measured in one run, within 10 hours of the cosmic muon event accumulation.

The one-square-meter ED consists of 16 plastic scintillation blocks, each with the size of $25cm\times25cm\times1.5cm$ and 8 light fibers to collect photons to PMT(photomultiplier tube). After the first prototype of ED is produced, it is tested with CORARS. The time resolution, detection efficiency, and photon yield and their uniformity are measured and compared to the simulation, which shows that the prototype has reached the requirement of design.

In section 2 we describe the CORARS system, including the structure, the electronics setup, the DAQ (Data Acquisition) system and the system calibration procedure. Section 3 is devoted to it's application in the test of the prototype of the ED and the results of the measurement are given in section 4.

\section{Description of CORARS}

\label{}
\subsection{The detector system}
\begin{figure*}[!hbt]
  \begin{minipage}{0.45\textwidth}
  \centering
    \includegraphics[width=0.8\columnwidth,height=8cm, viewport=0 0 490 800,clip,scale=0.4]{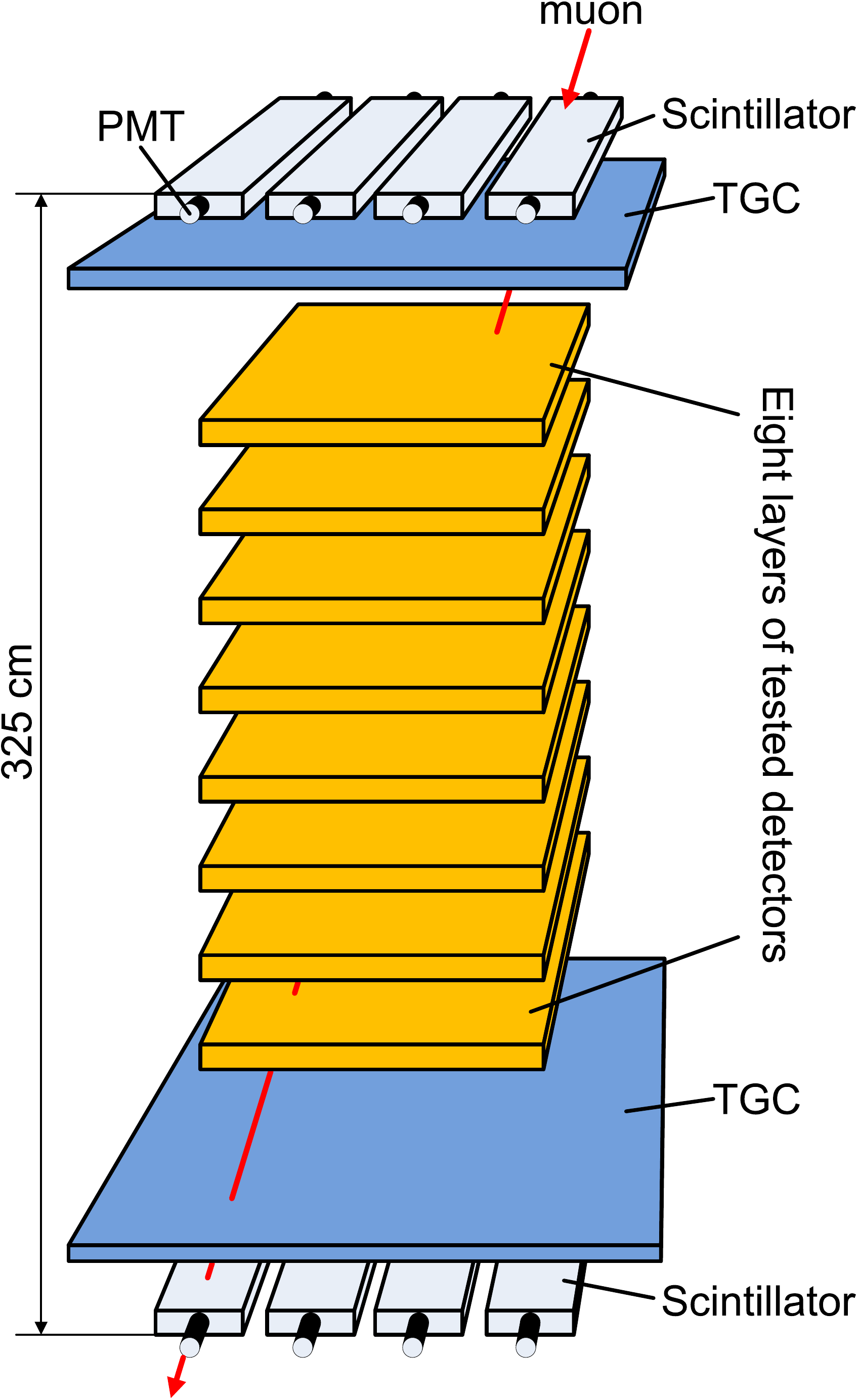}
  \end{minipage}
  \hspace{3ex}
  \begin{minipage}{0.45\textwidth}
    \centering
    \includegraphics[width=0.6\textwidth,viewport=0 0 455 830,clip,scale=0.4,angle=-90]{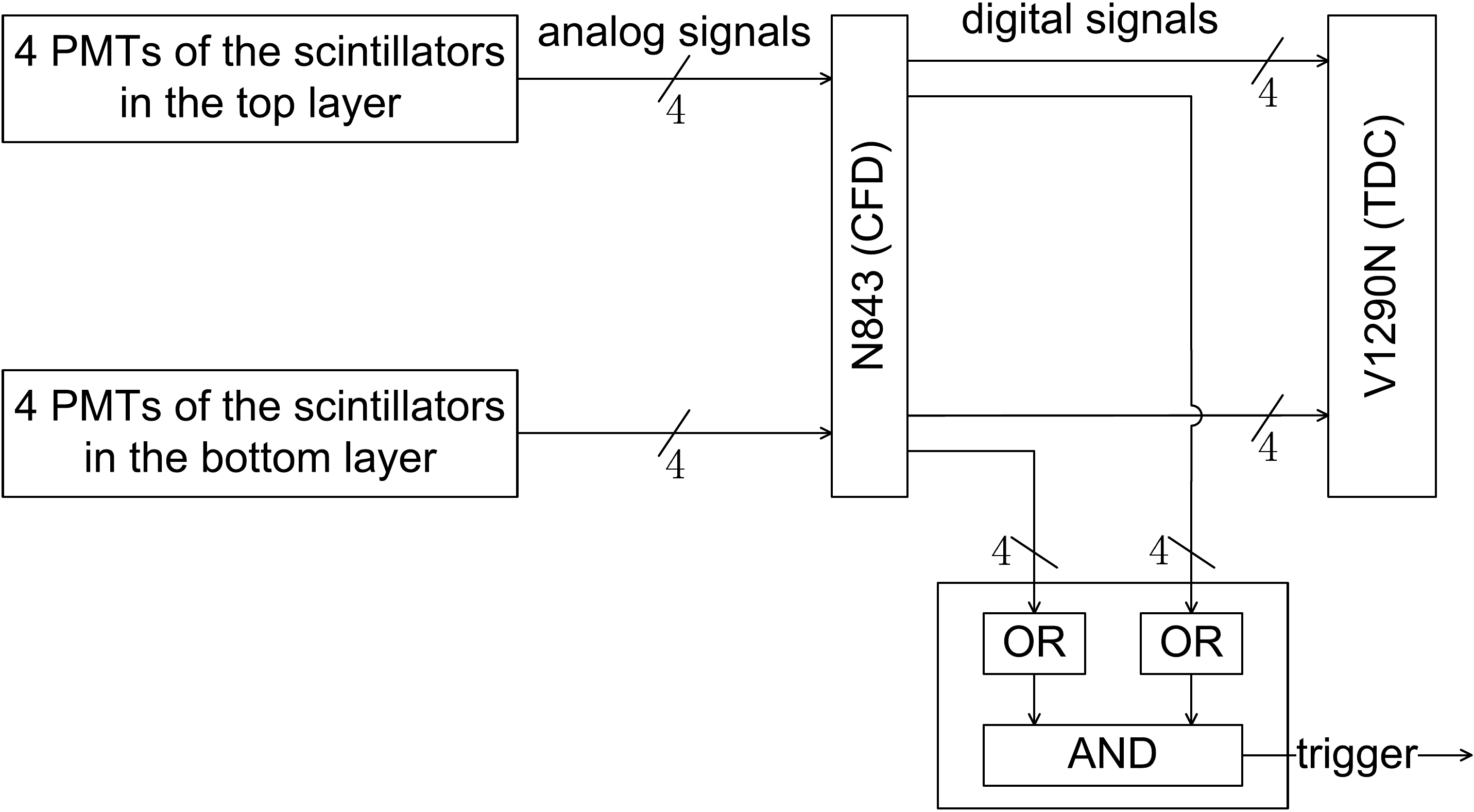}
  \end{minipage}
  \centering
  \caption{\small\sf (color online) Schematic layout of the detectors included in CoRaRS (left), and the processing of the PMT signals of the eight scintillator detectors (right).}
  \label{fig:detectors}
\end{figure*}

CORARS is $325cm$ high and $120cm\times120cm$ in intersecting plane as shown in Fig.~\ref{fig:detectors}. Four plastic scintillation detectors in rectangular shape of $1.2m\times25cm\times7cm$, are placed on the top and bottom most layer for system trigger generation and muon hitting time measurement. Thin Gap Chamber (TGC)~\citep{ATLAS1997ad}, in shape of isosceles trapezium, with bases of $1.5m$ and $1.7m$, height of $1.2m$, is placed on the layer adjacent to each scintillation detector layer for position measurement with a resolution of around $1cm$. The volume between the two TGC layers, are divided into layers to place tested detectors. By design, maximum eight EDs can be inserted for test simultaneously.

\subsection{Electronics and DAQ} \label{subsec:principleAndTrigger}
The analog signals from the PMT of the scintillation detectors are discriminated by Constant Fraction Discriminator (CFD). The NIM signal from the discriminator is then fed to a Time Digital Convertor (TDC) for timing measurement. The 4 signals from the scintillation detectors on the same layer are fed to the "or" logic and the two output from the "or" logic corresponding to the two scintillation detector layers are used to produce the trigger with a "and" logic, as shown in Fig.~\ref{fig:detectors}. By fine tuning the widths of the output NIM signals from CFD to be minimum($50ns$), the trigger generation is constrained by the time difference between the two "or" logic output to be as smaller as possible,  which can effectively reduce the fake triggers from noises.

Two-dimensional spacial information of the incident muon is given by TGC layers. On one TGC layer, 32 electronic channels are responsible for X-dimention measurement and 32 channels for the Y-dimension measurement. The LVDS signals from the two TGCs' frond-end electronic boards are recognized and recorded by a 128-channel TDC module.

Once the system is triggered, the signals from scintillation detectors and TGCs are read out, which happens around 10 Hz in the altitude of ~50 meters. By reconstruct the muon tracks as straight lines, the hitting time and position on the testing layers are calculated.

As the detectors to be tested, the time and charge of ED's signals are measured. The ED signal is firstly fanned out into two identical analog signals and one is sent to a charge-to-digital convertor (QDC), with LSB of $100fC$, with a proper delay, another is sent to CFD and then TDC for time measurement.

A software has been developed using \texttt{c\#} to collect and decode the raw data from the VME modules and then save in ROOT format. This software also does some rudimentary data analysis online to monitor the operation status of the system.


In order to eliminate the scintillating light propagation time offset in the scintillation materials due to the different hitting position, a timing calibration procedure was implemented. The sensitive area of the scintillation detector are divided into many pixels of size $2.5cm\times2.5cm$, considering the spacial sensitivity of the system. The mean time delay of every pixel relative to a global time reference was obtained. As the result of the one-side readout scintillation detector, the time delay increases along with the distance between the hit position and the PMT.  The width $\sigma_{delay}$ of the distribution of the time delay of each pixel serves as a figure-of-merit for the time resolution of that scintillator pixel. The $\sigma_{delay}$ of most of the pixels($92\%$) are smaller than $700ps$. The pixels whose $\sigma_{delay}$ are larger than $700ps$, are mostly around the end side where the PMT is mounted. In order to keep a better time resolution, these pixels were not used.

\begin{figure*}[!hbt]
  \begin{minipage}{\columnwidth}
    \includegraphics[viewport=35 15 560 260,clip,scale=0.4]{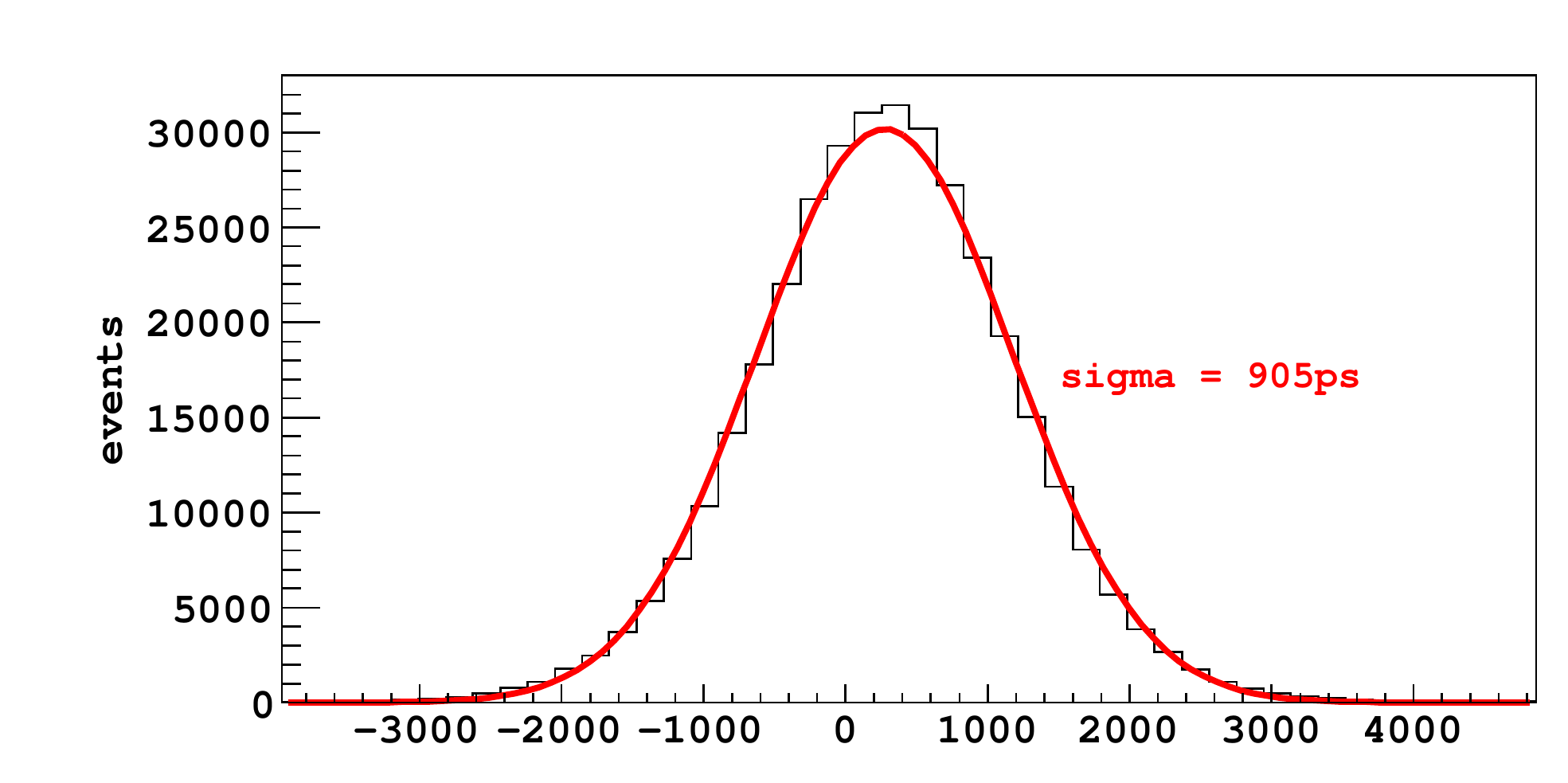}
    \centering
    \makebox{\scriptsize{($tof_{scin}-tof_{dis}$) (ps)}}
  \end{minipage}
  \hspace{3ex}
  \begin{minipage}{\columnwidth}
    \centering
    \includegraphics[viewport=25 15 560 290,clip,scale=0.37]{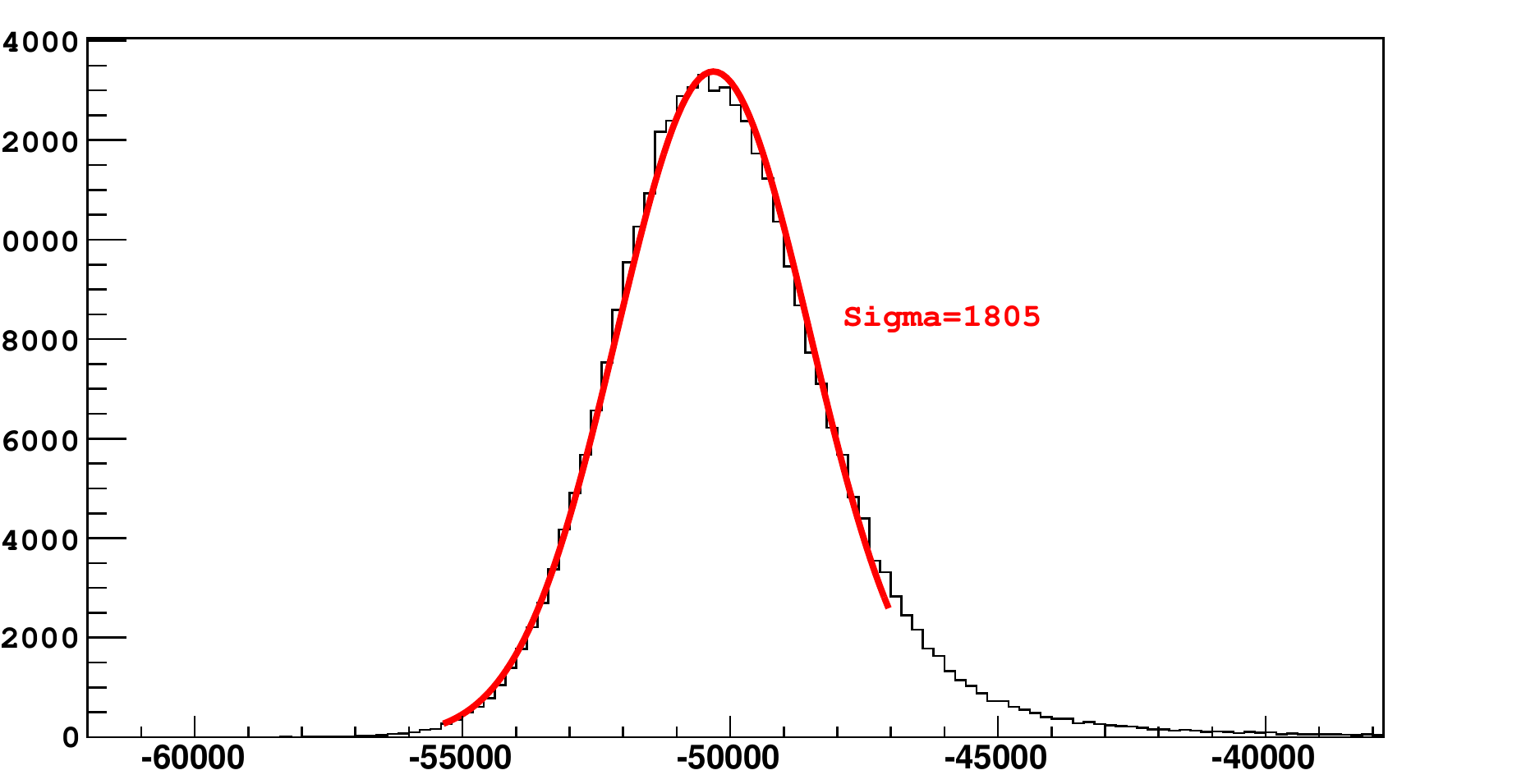}
    \centering
    \makebox{\scriptsize{($t_{ed}-t_{ed}^{reconstructed}$) (ps)}}
  \end{minipage}
  \hspace{3ex}
  \caption{\small\sf (color online) The distribution of the differences between the two time of flight obtained using two methods (left), and the distribution of the differences between the time given by ED and the reconstructed time by CORARS.(right).}
  \label{fig:sysSigma}
\end{figure*}

A test is then done to find out the calibrated time resolution of each layer of scintillation detectors. For a muon which passes through the entire detector system, the time of flight of that muon can be obtained using
\begin{equation} \label{equ:tofScin}
tof_{scin}=(t_{btm}-t_{top}),
\end{equation}
where $t_{btm}$ is the time when the muon hits the bottom scintillation detector layer, and $t_{top}$ for the top layer. The time of flight can also be obtained using
\begin{equation} \label{equ:tofDis}
tof_{dis}=d/v,
\end{equation}
where $d$ is the distance that the muon travels between the top and bottom scintillator layers, and $v$ is the velocity of the muon, which is assumed to be the same as the speed of light. Fig.~\ref{fig:sysSigma} shows the distribution of
\begin{equation}
\Delta tof=tof_{scin}-tof_{dis},
\end{equation}
and the width $\sigma_{\Delta tof}$ of this distribution is $905ps$, which is contributed mainly by the time resolutions of the two scintillator layers. The resolutions of the two scintillator layers are supposed to be the same, so the time resolution of each scintillator layer after calibration is
\begin{equation} \label{equ:sigmaScin}
\sigma _{scin}=\sigma _{\Delta tof}/\sqrt 2\approx640ps.
\end{equation}

\section{Test of ED}
The performances of the first ED prototype was studied using this system. The method and the results are presented.
\subsection{Time resolution}
Using the time $t_{top}^{initial}$ given by the top scintillator layer, $t_{btm}^{initial}$ given by the bottom scintillator, and the calibration constants of the scintillator detectors, the corrected hitting time on the system
\begin{equation}\label{equ:time}
t_{top}^{corrected}=t_{top}^{initial}-t_{top}^{delay}
\end{equation}
\begin{equation}\label{equ:time}
t_{btm}^{corrected}=t_{btm}^{initial}-t_{btm}^{delay}
\end{equation}
 can be obtained. Then the time when the ED is hit can be reconstructed using the formula
\begin{equation}\label{equ:timeReconstructed}
t_{ed}^{reconstructed}=t_{top}^{corrected}+\frac{h}{H}\cdot(t_{btm}^{corrected}-t_{top}^{corrected}),
\end{equation}
where $H=325cm$ is the vertical distance between the top and the bottom scintillator layers, and $h=224.25cm$ is the vertical distance between the top scintillator layer and the layer where the prototype ED is placed in the real test. On the other hand, ED also gives the time $t_{ed}$ when it is hit by the muon. The time resolution of the ED can be then obtained from the width $\sigma_{\Delta}$ of the distribution of
\begin{equation} \label{equ:timeDiff}
\Delta=t_{ed}-t_{ed}^{reconstructed}.
\end{equation}
The histogram of this distribution is shown in Fig.~\ref{fig:sysSigma}, and $\sigma_{\Delta}=1805ps$ is obtained, which are contributed from both the time resolutions of the ED and CORARS system. According to Eq.~\ref{equ:timeReconstructed}, the time resolution $\sigma_{t_{ed}^{reconstructed}}$ is given by
\begin{equation} \label{equ:sigmaTimeRecon}
\sigma_{t_{ed}^{reconstructed}}=\frac{\sqrt{(H-h)^{2}+h^{2}}}{H}\cdot\sigma_{scin}\approx487ps,
\end{equation}
Then the time resolution of the ED is obtained
\begin{equation} \label{equ:sigmaED}
\sigma_{ed}=\sqrt{(\sigma_{\Delta})^{2}-(\sigma_{ed}^{reconstructed})^{2}}\approx1740ps.
\end{equation}



In order to understand in more detail the uncertainties contributed to the time resolution of the whole ED, the uniformity of the time walking and time resolution is scanned by dividing the ED area into $5cm\times5cm$ pixels. The time resolution of each pixel is clearly less than 2ns, but the time walking of each pixel, which is shown in Fig.~\ref{fig:timeResUniformity}, is not perfect. The calibration of the PMT transition time relative to photon injection position on the PMT window shows that transition time is shorter in the center of the PMT window and longer on the edge. This prompt us that the non-uniformity of the transition time of PMT should be measured before used for the ED assembling.
\subsection{Photoyield and detection efficiency.}
\begin{figure*}[!hbt]
  \begin{minipage}{0.45\textwidth}
  \centering
  \includegraphics[viewport=30 10 550 270,clip,width=1\textwidth]{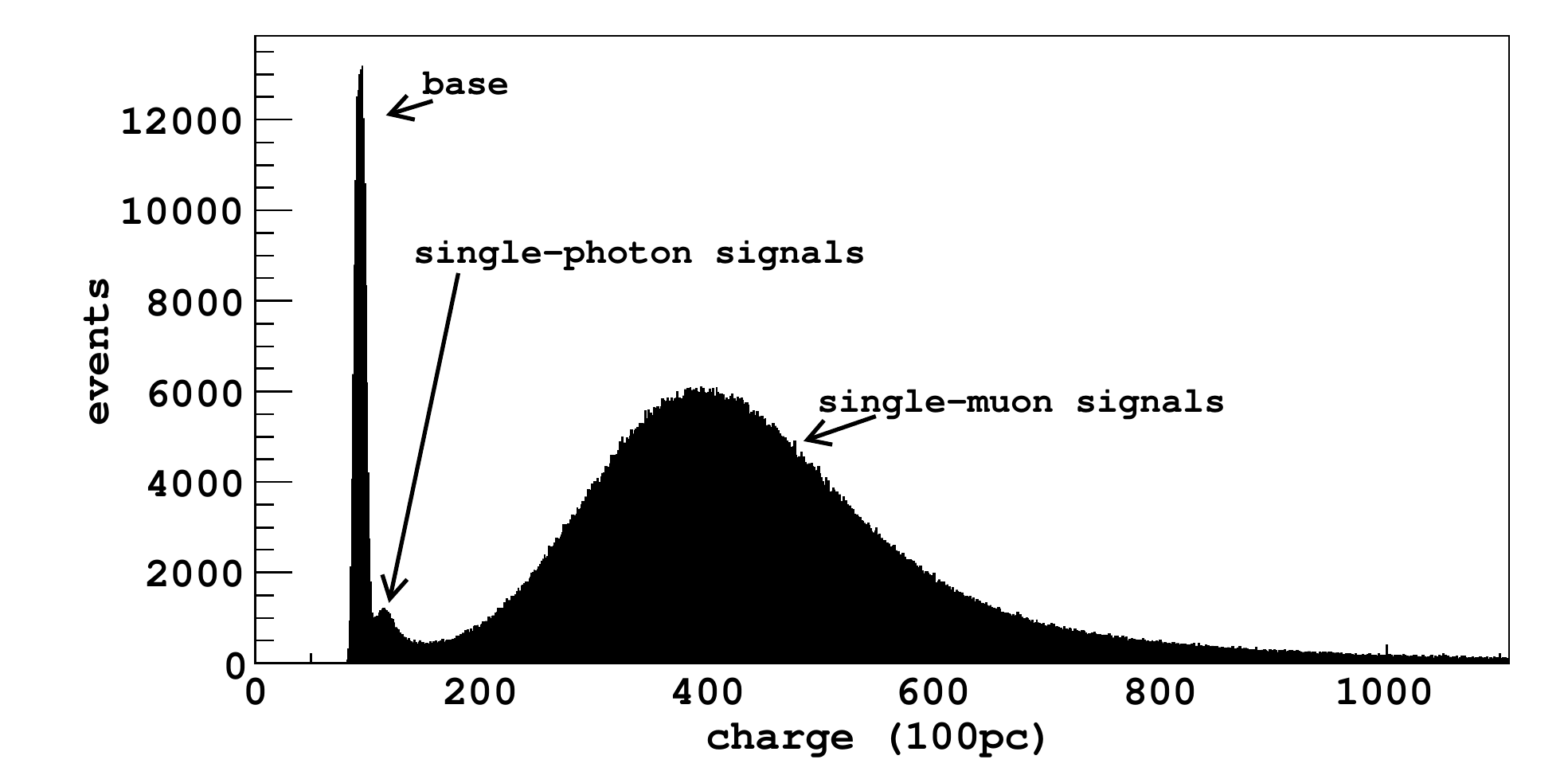}
  \end{minipage}
  \hspace{3ex}
  \begin{minipage}{0.45\textwidth}
  \centering
  \includegraphics[viewport=10 35 543 475,clip,width=1\textwidth]{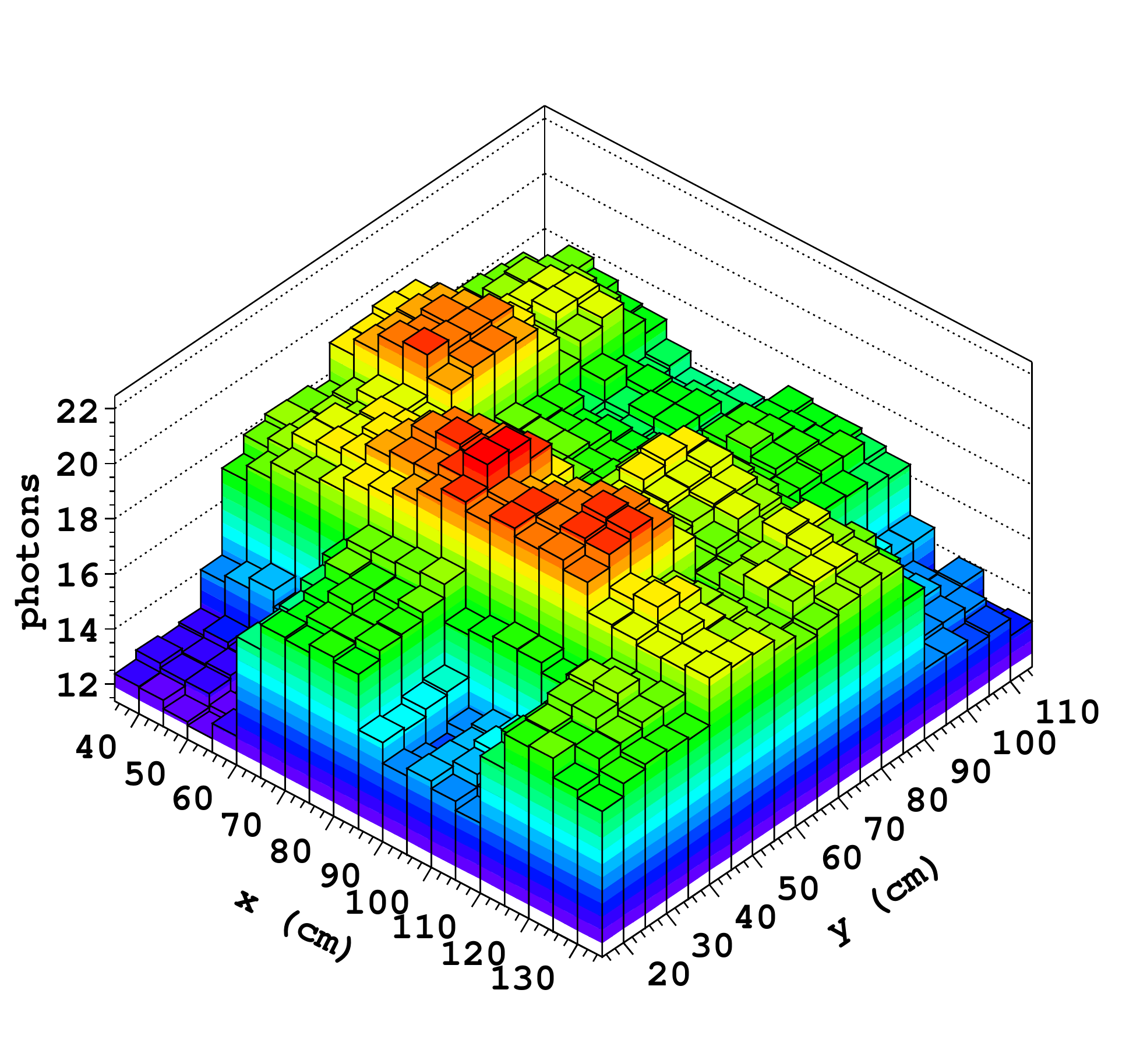}
  \end{minipage}
  \caption{\small\sf (color online) The charge spectrum of the ED signals (left), and the uniformity of the average photo-yield converted from the charge of ED signal (right).}
  \label{fig:q}
\end{figure*}
In this test, the charges of the signals from ED are measured. In the charge spectrum showed in the left part of Fig.~\ref{fig:q}, three peaks can be seen. The first peak is the electronic pedestal. The second peak is the charges of single-photon signals, and the last one is the charges of single-muon signals. The mean charge of pedestal is $9.3pc$, the mean charge of single-photon signals is $11.4pc$, and the average charge of single-muon signals is $46.6pc$. It can be calculated that $17.8$ photon-electrons on average are collected by PMT when one cosmic muon passes through ED, i.e., the average photon-yield is $17.8 photon/muon$. The photon-yield of each $5cm\times5cm$ pixel of ED was calculated and is shown in the right part of Fig.~\ref{fig:q}, which shows the photon-electron collection efficiency between scintillation blocks are some different, which can be explained by the non-uniformity of the PMT quantum efficiency over the whole PMT window. The scintillation block with lowest photon-yield is known with one light fiber broken.

\begin{figure*}[!hbt]
  \begin{minipage}{0.55\textwidth}
  \centering
  \includegraphics[viewport=15 20 410 310,clip,scale=0.65]{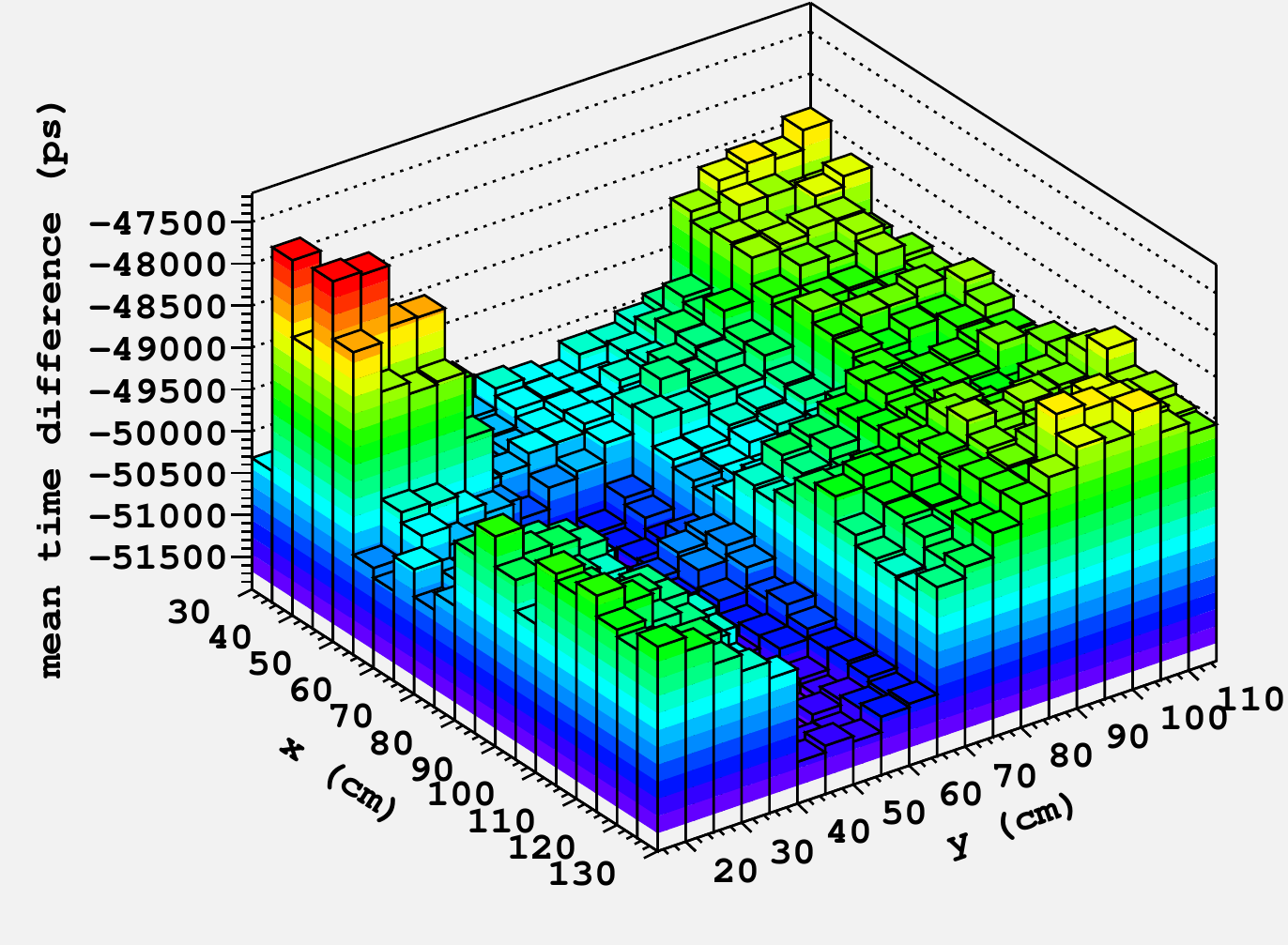}
  \end{minipage}
  \hspace{3ex}
  \begin{minipage}{0.45\textwidth}
  \centering
  \includegraphics[viewport=15 60 560 500,clip,scale=0.45]{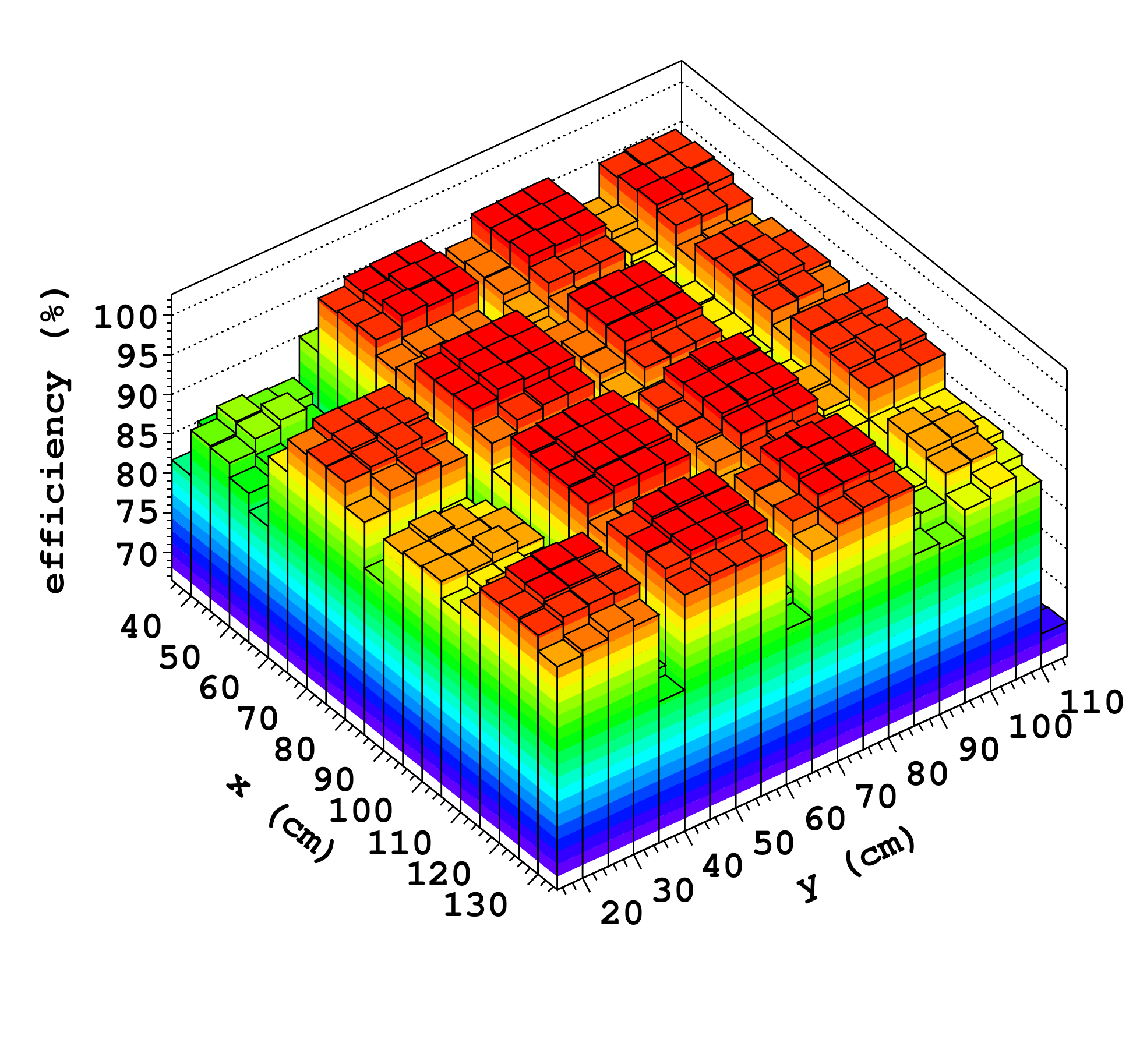}
  \end{minipage}
  \caption{\small\sf (color online) The uniformity of the time walking (left) relative to trigger and the the detection efficiency (right) of each $5cm\times5cm$ pixel of ED. }
  \label{fig:timeResUniformity}
\end{figure*}

When a cosmic muon is confirmed by CORARS to hit the ED, the signal from ED is checked. If ED signal presents, this muon is considered as being detected by ED, otherwise it is considered as being missed by ED. The radio of the number of events being detected by ED to the number of total events is the detection efficiency of the ED. A average detection efficiency of about $93.7\%$ was obtained for the entire ED. The detection efficiency of each $5cm\times5cm$ pixel was also calculated, and the result is showed in Fig.~\ref{fig:timeResUniformity}. This ``lower'' efficiency is partially due to threshold definition of the electronic system, and can be adjusted later.
\section{Summary}
 A CORARS system, which has good time ($~0.5ns$) and spacial (~1cm) resolution, is built. It's obvious that the scanning ability of CORARS is crucial in the ED testing and the problem finding, which is desired in the quality control of a big quantity of detectors. The first ED prototype for LHAASO project is tested, showing a good performance as designed. The time resolution is measured to be around $1.7ns$, which meets the requirement of the design. About 17.8 photon-electrons are collected on average by PMT when a cosmic muon passes through the ED, which is consistent with the simulation. The mean detection efficiency of the entire ED is about $93.7\%$, which suffer from the worse performance of 1 scintillation block with light fiber broken. \\








\end{multicols}

\clearpage
\end{CJK*}

\end{document}